\documentclass[prb,twocolumn,showpacs,floatfix,amsfonts]{revtex4}
\usepackage{graphicx,graphics,color,epsfig}
\usepackage{bm}
\usepackage{amsmath}
\usepackage{amssymb}

\begin{document}
\preprint{}

\title{Local density of states in a $d$-wave superconductor with 
stripe-like modulations and a strong impurity}
\author{Hong-Yi Chen and C. S. Ting}
\affiliation{Texas Center for Superconductivity and Department of Physics,
University of Houston, Houston, TX 77204}

\begin{abstract}
Using an effective Hamiltonian with $d$-wave superconductivity (dSC) and  
competing antiferromagnetic (AF) interactions, we show that weak and  
one-dimensionally modulated dSC, spin density wave (SDW) and charge  
density wave (CDW) could coexist in  the ground state configuration.
With proper parameters, the SDW order exhibits a period of 8a, while for 
dSC and CDW orders the period is 4a.
The local density of states (LDOS), which probing the behavior of 
quasiparticle excitations, is found to have the identical stripe-like 
structure as those in dSC and CDW orders. The LDOS as a function of the 
bias voltage are showing two small bumps within the superconducting 
coherence peaks, a signature of the presence of stripes. When a strong 
impurity like Zn is placed in such a system, the LDOS at its nearest 
neighboring sites are suppressed at the zero bias by the local AF order 
and show a double-peak structure.
\end{abstract}


\maketitle

Many of the anomalous properties of high-$T_c$ superconductors (HTS) are 
believed to be related to the competition between the $d$-wave  
superconductivity (dSC) and the hidden antiferromagnetic (AF) order, 
particularly in the underdoped HTS .
Inelastic neutron scattering (INS) studies of the magnetic fluctuations 
in some of the HTS samples have provided 
important clues to the nature of the electronic correlations within the  
doped CuO$_2$ planes.\cite{yamada98,tranquada95,mook00}.
In additon, the existence of one-dimensional charge density wave (CDW) 
and spin density wave (SDW) has also been reported on 
La$_{1.6-x}$Nd$_{0.4}$Sr$_x$CuO$_4$ \cite{tranquada97} and 
YBa$_2$Cu$_3$O$_{6.35}$ \cite{mook02}. 

The elastic neutron scattering measurements on
La$_{2-x}$Sr$_x$CuO$_4$ with $x=0.10$ sample by Lake $et\;al.$ 
\cite{lake415} found that the signal got enhanced at 
$(\frac{1}{2},\frac{1}{2}+\delta)(\frac{2\pi}{a})$
at low temperature, indicating the presence of static AF order at least in 
some parts of the sample. 
The coexistence of the CDW and dSC in HTS in the absence of a magnetic 
field has attracted a lot of theoretical works
\cite{salkola77,martin14,ychen66,ichioka02,podolsky67}.
Recent progress in scanning tunneling microscopy (STM) on the surface of  
Bi$_2$Sr$_2$CaCu$_2$O$_{8+\delta}$ (BSCCO) has given us a high resolution 
probe of its electronic correlations. The STM imaging by Hoffman 
$et\;al.$ \cite{hoffman02} in the presence of a magnetic field
revealed the quasiparticle states around the vortex cores in slightly 
overdoped BSCCO to exhibit a CDW of "checkerboard" pattern with $4a$ 
periodicity. Howald $et\;al.$ \cite{howald0201,howald0208} reported that 
the four unit-lattice periodic charge modulation survives even at zero 
magnetic field and is energy independent. And it has been attributed 
to the superposition of stripe phases \cite{howald0201,kivelson0210} 
oriented along $x$- and $y$- directions in the sample.
This issue has also been examined by several theoretical articles 
in the presence of a magnetic field
\cite{ychen66,yzhang66,zhu89,hdchen89,polkovnikov65}
and in the absence of a magnetic field \cite{podolsky67}.

In case a stripe phase indeed exists in some of the HTS samples close to 
optimal doping as reported in the experiments  
\cite{howald0201,howald0208}, it should have important impact on the 
physics of HTS. We need to understand the origin of the stripe-like 
modulation observed in the local density of states (LDOS) and to examine 
its consequence  on other experimentally measurable quantities, such as 
the LDOS near a unitary impurity in the absence of a magnetic field. In 
view of neutron scattering experiments mentioned above, we assume that 
the stripe phase observed in the STM experiments is originated in the 
competing AF interaction. In this paper and based upon a model 
Hamiltonian, we first construct a superconducting phase for a nearly 
optimal-doped or slightly underdoed HTS sample in which the dSC coexists 
with the SDW, and CDW orders.
The parameters will be chosen in such a way that the dSC and CDW will have 
a weak one-dimension-like modulation with period 4a while the period for 
the SDW is 8a.
The superposition of the $x$- and $y$- oriented CDW stripes should 
yield the experimentally observed checkerboard patterns 
\cite{howald0201,howald0208}.
Then we calculate the LDOS of the stripes and compare it with the pure 
$d$-wave case. 
On the other hand, the low temperature STM experiment \cite{pan403} 
observed a sharp resonance peak near the zero bias at a nonmagnetic 
unitary impurity site, consistent with prediction \cite{balatsky51}
for a pure dSC.
In the presence of the stripe phase, we found that the LDOS is strongly 
suppressed by the AF order at the zero bias and a double-peak shows up.
Hopefully, the obtained result could be used to compare with future STM 
experiments for samples with magnetic originated stripes.   

Based on a square lattice with lattice constant a(=1),we start from the 
effective mean-field Hamiltonian with $g$ as the nearest neighbor 
attractive $d$-wave pairing potential, and $U$ as the on-site Coulomb 
interaction representing the competing AF order.

\begin{eqnarray}
H &=&-\sum_{{\bf ij},\sigma} t_{\bf ij} c_{{\bf
i}\sigma}^{\dagger}c_{{\bf j}\sigma} 
+\sum_{\bf ij} (\Delta_{\bf ij} c_{{\bf i}\uparrow}^{\dagger}
c_{{\bf j}\downarrow}^{\dagger} +\Delta_{\bf ij}^{*} c_{{\bf
j}\downarrow} c_{{\bf i}\uparrow} ) \nonumber \\
&& + \sum_{\bf i} (m_{{\bf i}\sigma}-\mu+V_{\bf i}) c_{{\bf
i}\sigma}^{\dagger}c_{{\bf i}\sigma}
\end{eqnarray}
where $t_{\bf ij}$ is the hopping integral,
$\Delta_{\bf ij} = \frac{g}{2}\langle c_{{\bf i}\uparrow} c_{{\bf
j}\downarrow}-c_{{\bf i}\downarrow}c_{{\bf j}\uparrow}\rangle $ 
is the spin-singlet $d$-wave bond order parameter,
$m_{{\bf i}\sigma}=U\langle n_{{\bf i}\sigma}\rangle$ is the AF order 
parameter,
$\mu$ is the chemical potential.
In addition there is also a nonmagnetic impurity with $V_{\bf 
i}=V_0\delta_{0\bf i}$ as the single-site scattering potential at site 
$(0,0)$.
We shall diagonalize the above Hamiltonian by using Bogoliubov 
transformation, $c_{{\bf i}\sigma} = \sum^N_n [ u_{{\bf i}\sigma}^n 
\gamma_{n\sigma} - \sigma v_{{\bf i}\sigma}^{n*} 
\gamma_{n\bar{\sigma}}^\dagger ]\;$, 
and the equations of motion for $c_{{\bf i}\sigma}$ and $c_{{\bf 
i}\bar{\sigma}}^\dagger$ will lead to usual Bogoliubov-de Gennes' 
equations (BdG),
\begin{eqnarray}
\sum_{\bf j}^N  \left(\begin{array}{cc}
                   {\cal H}_{{\bf ij}\sigma} & \Delta_{\bf ij} \\ 
                   \Delta_{\bf ij}^* & -{\cal H}_{{\bf ij}\bar{\sigma}}^*
                \end{array}\right)
                \left(\begin{array}{c}
                   u_{{\bf j}\sigma}^n \\
                   v_{{\bf j}\bar{\sigma}}^n
                \end{array}\right)
                = E_n
                \left(\begin{array}{c}
                   u_{{\bf i}\sigma}^n \\ 
                   v_{{\bf i}\bar{\sigma}}^n 
                \end{array}\right)\;,
\end{eqnarray}
where
${\cal H}_{{\bf ij}\sigma}=-t \delta_{{\bf i}+\hat{\bf
e},{\bf j}} + (m_{{\bf i}\bar{\sigma}}-\mu) \delta_{\bf
ij}+V_{\bf 0}\delta_{\bf 0j}$.
The subscript $\hat{\bf e}$ are the vectors $\hat{x}$, $\hat{y}$ and 
$\hat{x}+\hat{y}$ toward the nearest neighbor (NN) and next-nearest 
neighbor (NNN) sites, respectively.
To self-consistently solve BdG equations we could get the $N$ positive 
eigenvalues $E_n$ $(n=1\cdots N)$ and the $N$ negative eigenvalues 
$\bar{E}_n$ with corresponding eigenvectors $(u_{{\bf i}\uparrow}^n , 
v_{{\bf i}\downarrow}^n)$ and $(-v_{{\bf i}\uparrow}^{n*} , u_{{\bf 
i}\downarrow}^{n*} )$, respectively. Using the following convenient 
notation
$\bar{u}_{{\bf i}\uparrow}^n = ( -v_{{\bf i}\uparrow}^{n*},
        u_{{\bf i}\uparrow}^n )$ and 
$\bar{v}_{{\bf i}\downarrow}^n = ( u_{{\bf i}\downarrow}^{n*}, 
	v_{{\bf i}\downarrow}^n )$,
the self-consistent conditions become,
\begin{eqnarray}
\langle n_{{\bf i}\uparrow} \rangle &=&
        \sum_{n=1}^{2N}\left|\bar{u}_{{\bf i}\uparrow}^n\right|^2f(E_n)
	\;,\;
\langle n_{{\bf i}\downarrow} \rangle =
        \sum_{n=1}^{2N}\left|\bar{v}_{{\bf i}\downarrow}^n\right|^2 
        [1-f(E_n)] \nonumber \\
\Delta_{\bf ij} &=& \sum_{n=1}^{2N} \frac{g}{4}
(\bar{u}_{{\bf i}\uparrow}^n \bar{v}_{{\bf j}\downarrow}^{n*} + 
\bar{v}_{{\bf i}\downarrow}^{n*} \bar{u}_{{\bf j}\uparrow}^n) \tanh 
(\frac{\beta E_n}{2}) \;,
\end{eqnarray}
where $f(E)=1\slash(e^{\beta E}+1)$ is Fermi distribution function.

Since the calculation will be performed near the optimal-doped or 
slightly underdoped regime, we choose the filling factor, which is defined
as $n_f=\sum_{{\bf i}\sigma} \langle c_{{\bf i}\sigma}^\dagger c_{{\bf 
i}\sigma} \rangle \slash N_x N_y$ with the summation over one unit cell, 
is fixed to be 0.85 (i.e. the hole doping $x_h=0.15$), where $N_x$, $N_y$ 
are the linear dimension of the unit cell under consideration. The 
chemical potential $\mu$, therefore, needs to be adjusted each time when 
the on-site repulsion $U$ is varied. 
The NNN hopping integral is chosen to be $t'=-0.2$, as relevant to the 
hole-doped cuprate, to fit the hole-like Fermi surface.
Through out this paper, we are going to use the same value of $U$, $g$, 
$t'$, and $n_f$.

Once the self-consistent solution is obtained, we calculate the staggered 
magnetization and the electron density defined as
$M_s=(-1)^{\bf i}\langle n_{{\bf i}\uparrow}-n_{{\bf i}\downarrow}\rangle$ 
and
$n_{\bf i}=\langle n_{{\bf i}\uparrow}+n_{{\bf i}\downarrow}\rangle$,
respectively.
Furthermore, the LDOS of the energy $E$ at the positon ${\bf i}$ can be 
written as 
\begin{eqnarray}
\rho_{\bf i}(E) &=&-\frac{1}{M_x M_y} \sum_{n,{\bf k}}^{2N}
        \left| \bar{u}_{{\bf i}\uparrow}^{n,{\bf k}} \right|^2 
        f'(E_{n,{\bf k}}-E) \nonumber \\
        &&+\left| \bar{v}_{{\bf i}\downarrow}^{n,{\bf k}} \right|^2 
        f'(E_{n,{\bf k}} + E)\;,
\end{eqnarray}
where $\rho_{\bf i}(E)$ is proportional to the local differential 
tunneling conductance as measured by STM experiment, and the summation is 
averaged over a $M_x \times M_y$ wavevectors in first Brillouin Zone.
In addition, the LDOS spatial maps observed in STM experiment at a fixed 
bias voltage $E$ could be obtained by calculating $\rho_{\bf i}(E)$ at 
each site of the lattice.

In the absence of the magnetic field, $t_{\bf ij}=t$ and $t'$ are the 
NN and NNN hopping integrals, respectively.
Since the strength of the onsite repulsion, $U$, required to generate AF 
order has to be fine tuned to study the interplay of the two competing 
orders.
The non-zero AF order appears only for certain range of $U$ and $g$, and 
larger $U/g$ is able to produce stronger antiferromagnetism.
With our chosen parameters $U=2.05$ and $g=0.71$, weak stripe modulations 
can be generated in the dSC order parameter, the staggered magnetization 
and electron density. Here we set $t=1$.

\begin{figure}[t]
\centerline{\epsfxsize=8.5cm\epsfbox{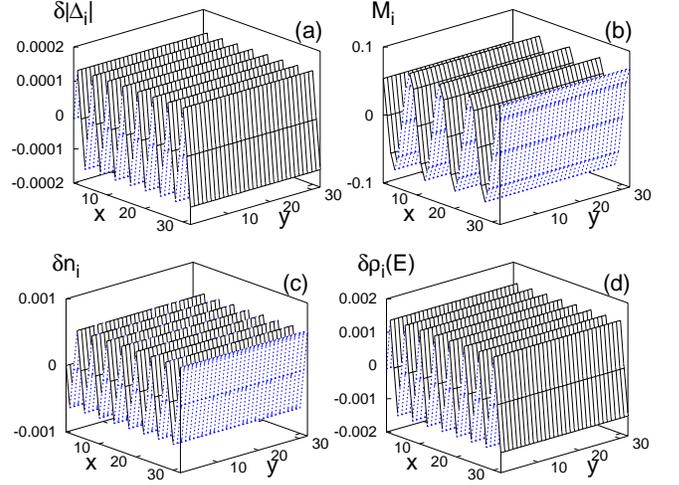}}
\caption[*]{
The spatial distribution of 
(a) The dSC order parameter $\delta|\Delta_{\bf i}|=|\Delta_{\bf 
i}|-0.041$, 
(b) The staggered magnetization ,and
(c) the electron density $\delta n_{\bf i} = \sum_\sigma n_{{\bf i}\sigma} 
- n_f$ for the pure superconductor (without impurity). 
(d) The LDOS maps $\delta\rho_{\bf i}(E)=\rho_{\bf i}(E)-0.084$ at the 
energy $E=0.04$.
The size of the unit cell is $N_x\times N_y = 32\times 32$. The attractive 
potential is $g=0.71$. The on-site repulsion is $U=2.05$. The NNN hopping 
term is $t'=-0.2$. The filling factor is $n_f=0.85$.
}
\end{figure}

In Figure 1(a), our numerical result for the spatial distribution of the 
dSC order, which is defined as
$\Delta_{\bf i} = [\Delta_{{\bf i}+\hat{x}} +
\Delta_{{\bf i}-\hat{x}} - \Delta_{{\bf i}+\hat{y}} - 
\Delta_{{\bf i}-\hat{y}}] / 4\;,$
exhibits the stripe behavior along $y$-direction with the period $4a$. 
The staggered magnetization (Fig. 1(b)) shows stripe like SDW along 
$y$-direction. Its period is $8a$ and its amplitude is less than $0.1$.
The electron density (Fig. 1(c)) has one-dimensional CDW modulation with 
$4a$ as its period and a very weak amplitude (less than $0.001$).
The origin of such static stripes could be understood in terms of the  
existence of a nesting wave vector $q_A \sim 0.25\pi/a$ connecting the 
upper and lower pieces of the Fermi surface near $(\pi,0)$ along 
$k_y$-direction
[See Fig. 1 in Refs. 24]. For proper values of $U$ and doping, this wave
vector would modulate the staggered magnetization with SDW stripes along
$x$-direction with period $2\pi/q_A=8a$. Accompanying the SDW stripes 
we have the charge stripes with period $4a$. If we choose a larger 
$U(>2.05)$ here, the effect is to enhance the amplitudes of the SDW/CDW 
stripes, not to change its period. For smaller $U(=2.0)$, only dSC without 
stripes is obtained.
The stripe configurations discussed above are associated with the ground 
state of the system. What have been observed by STM experiments are 
related to quasiparticle excitations. In Fig. 1(d) we present the spatial 
map of the LDOS at energy $E=0.04$, and it exhibits the same stripe like 
modulation with period $4a$ as that appeared in CDW shown in Fig 1(c). We 
have checked the LDOS at several different bias energies, and found that 
the same stripe like structure still prevails and the modulation period 
or wavevector is energy independent. This result can be regarded as a 
necessary condition for existing a static-stripe CDW order in the sample.
Since the $y$- and $x$- oriented stripes are degenerated in energy. The 
observed checkerboard pattern in the STM experiments 
\cite{howald0201,howald0208} could be understood by a superposition of 
these two perpendicularly oriented LDOS maps \cite{ychen66}.
Part of the modulation wavevectors observed in the experiments 
\cite{hoffman297} are also dispersive or energy dependent, this 
behavior should be related to the scattering of quasiparticles from 
defects including the randomly distributed stripes which break the 
translational invariance \cite{qwang67,dzhang67}.
Here we would like to emphasize that our quasiparticles are excitations 
from the SDW/CDW (stripes) + dSC ground state. The information of the 
stripes is nonpertubatively included in wave functions of the 
quasiparticles. After the Hamiltonian is diagolized, there exists no extra 
term which couples the quasiparticle with the stripes.
It should also be interesting to calculate the energy-dependent LDOS in 
the presence of stripes and to see its difference as compared with the 
case for a pure dSC. In Fig. 2, we show the energy dependence of the LDOS 
at one of the hole (electron) accumulated stripes, namely at one of the 
minima (maxima) in Fig. 1(c).
It appears that there exist two small bumps within the superconducting 
coherence peaks, a signature of the presence of weak SDW and CDW orders. 
At the different sites, the LDOS also can be shown to have the similar 
features.

\begin{figure}[t]
\centerline{\epsfxsize=7.0cm\epsfbox{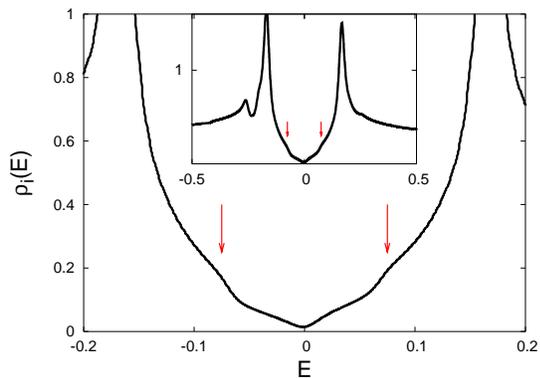}}
\caption[*]{
The LDOS as a function of energy at the hole (electron) accumulated site 
(stripe).
The parameter values are the same as the Fig. 1.}
\end{figure}

Next let us introduce a nonmagnetic strong impurity like Zn into our 
system at site $(0,0)$ and investigate its effect.
Here the on-site impurity potential is taken to be $V_0=100$. 
From Fig. 3(a), one can easily see that the dSC order is suppressed
and recovers to its bulk value at a length scale of $\xi_0$, the 
superconducting coherence length ($\sim 5a$), away from the impurity.
Beyond this range, the weak stripe-modulation in the dSC order parameter 
still remains.
In Fig. 3(b) we show that the impurity induced staggered-moment of the SDW 
is zero at the impurity site and reach the maximum value at its four NN 
sites.
The net induced local moment by the impurity corresponds to a local spin 
with $S_z$=1/2 when $U$ becomes stronger \cite{zwang89}.
From Fig. 3(b), the SDW is clearly pinned at the impurity site with one of 
its ridges.
The spatial profile of the electron density change is presented in 
Fig.3(c), and it exhibits a Freidel-like oscillation around the impurity. 
It is easy to see that the electron number reaches to zero or 
$\delta n_i=-0.85$ at the impurity site, which is much larger than the 
amplitude (~0.001) of the CDW stripes. This is the reason why the stripes 
in Fig. 1(c) are too faint to see in Fig.3(c). Fig. 3(d) displays the 
LDOS map at energy $E=0.04$. The intensity is 
zero at the impurity site and there are four peaks appear on the four NN 
sites around the impurity. The intensities of the two peaks at the 
site $(0,\pm1)$ are less than those at the site $(\pm1,0)$. The stripe 
structure in Fig. 1(d) still prevails, but it is again too weak to see 
here.

\begin{figure}[t]
\centerline{\epsfxsize=8.5cm\epsfbox{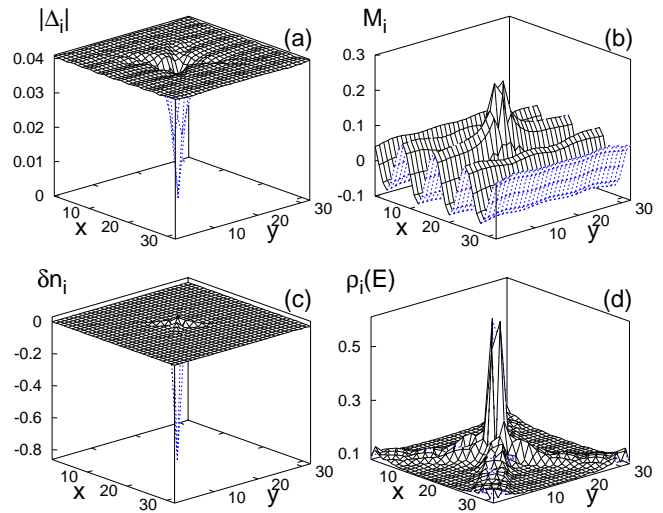}}
\caption[*]{
The spatial distribution of (a) the dSC order parameter, (b) the staggered 
magnetization ,and (c) the electron density variation  
with a single strong impurity at site $(0,0)$.
(d) The LDOS maps at the energy $E=0.04$.
The parameter values are the same as the Fig. 1, except the impurity 
scattering potential $V_0=100$.
}
\end{figure}

It is well known that a sharp single resonance peak at zero bias 
\cite{balatsky51} in the LDOS at the NN sites of a unitary impurity 
appears for a pure dSC.
In the presence of weak stripes, the LDOS as a function of energy at the 
NN site, $(0,1)$ and $(1,0)$ around the impurity are respectively 
displaced in Fig. 4(a) and (b).
The zero bias peak previously obtained for $U=0$ is dramatically 
depressed by $U\neq 0$ and splits into two distinct peaks.
This is because stronger local AF order is induced near the strong
impurity site and a larger SDW gap opens up locally that suppresses the
LDOS close to the impurity. This makes the LDOS in the present case very 
different from that of a pure d-wave (or $U=0$) case. The oscillation in
the LDOS at negative bias below the left peak is originated from the
energy-dependent modulations induced by the impurity, and the large
amplitude seems to come from the size effect.
When $U/g$ is not large enough to generate extended SDW, the weak and 
local AF order induced by the impurity may force the resonance peak just 
to split a little \cite{zwang89,tsuchiura64}.
The shapes of split peaks at site $(0,1)$ and $(1,0)$ are slightly 
different.
This indicates that the fourfold symmetry for a pure dSC changes to 
twofold symmetry when the stripe phase is in presence.
If the experimentally observed \cite{howald0201,howald0208} stripe 
structure is of magnetic origin, we predict that the LDOS should exhibit a 
double-peak feature near the zero bias.

\begin{figure}[t]
\centerline{\epsfxsize=6.0cm\epsfbox{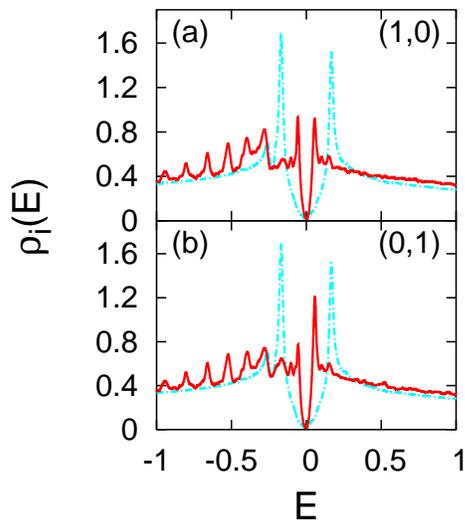}}
\caption[*]{
The LDOS as a funtion of energy at the NN site (a) $(0,1)$ 
and (b) $(1,0)$ around the single strong impurity located at the 
site $(0,0)$.
The parameter values are the same as the Fig. 1, and the impurity 
scattering potential $V_0=100$.
}
\end{figure}

In conclusion, we have studied a cuprate superconductor with weak 
stripe-like modulations 
in the dSC, SDW and CDW order parameters without and with  a strong 
nonmagnetic impurity.
In the absence of the impurity,
the LDOS exhibits two small bumps within the superconducting coherence 
peaks, a signature of the presence of the competing AF order.
The LDOS maps displays the same stripe modulation as the CDW along 
$y$-direction with periodicity $4a$ which is nondispersive or energy 
independent.
The components of the modulation wavevectors observed by the experiments 
which are energy dependent or dispersive should be attributed to 
the effect due to scatterings of quasiparticles from defects 
\cite{qwang67,dzhang67}.
In the presence of a strong impurity, we predict that the LDOS at the 
NN sites of the impurity are strongly suppressed by the AF order and 
reveal a double peak structure. Hopefully our theoretical results could be 
useful to future STM experiments performed on samples with weak 
stripe-like structures.

${\bf Acknowledgements}$: We thank Profs. S.H. Pan and N. C. Yeh for the 
useful discussions. This work is supported by The Texas Center for 
Superconductivity  at University of Houston and by a grant Robert A. Welch 
Foundation.

\end{document}